\documentclass[runningheads]{llncs}

\usepackage[hidelinks]{hyperref}
\usepackage[noabbrev,capitalize]{cleveref}
\usepackage{booktabs}
\usepackage{graphicx}
\usepackage{tabularx}
\usepackage{caption}
\usepackage{subcaption}
\usepackage{xcolor}
\usepackage[official]{eurosym}
\usepackage{lscape}
\usepackage{changepage}

\definecolor{xgreen}{HTML}{2bc253}
\definecolor{xyellow}{HTML}{fcba03}
\definecolor{xpink}{HTML}{f26884}

\DeclareCaptionType{Fig.}

\begin{document}

\title{A Co-design Study for Multi-Stakeholder Job Recommender System Explanations}
\titlerunning{Co-designing multi-stakeholder explanations}


\author{Roan Schellingerhout\orcidID{0000-0002-7388-309X} \and Francesco Barile\orcidID{0000-0003-4083-8222} \and Nava Tintarev\orcidID{0000-0002-5007-5161}}

\date{April 2023}

\institute{Department of Advanced Computing Sciences, Maastricht University, Maastricht, The Netherlands\\ \email{\{roan.schellingerhout,f.barile,n.tintarev\}}@maastrichtuniversity.nl}

\pagenumbering{arabic}

\authorrunning{R.P.F. Schellingerhout et al.}

\maketitle

\begin{abstract}
    Recent legislation proposals have significantly increased the demand for eXplainable Artificial Intelligence (XAI) in many businesses, especially in so-called `high-risk' domains, such as recruitment. Within recruitment, AI has become commonplace, mainly in the form of job recommender systems (JRSs), which try to match candidates to vacancies, and vice versa. However, common XAI techniques often fall short in this domain due to the different levels and types of expertise of the individuals involved, making explanations difficult to generalize. To determine the explanation preferences of the different stakeholder types - candidates, recruiters, and companies - we created and validated a semi-structured interview guide. Using grounded theory, we structurally analyzed the results of these interviews and found that different stakeholder types indeed have strongly differing explanation preferences. \textit{Candidates} indicated a preference for brief, textual explanations that allow them to quickly judge potential matches. On the other hand, \textit{hiring managers} preferred visual graph-based explanations that provide a more technical and comprehensive overview at a glance. \textit{Recruiters} found more exhaustive textual explanations preferable, as those provided them with more talking points to convince both parties of the match. Based on these findings, we describe guidelines on how to design an explanation interface that fulfills the requirements of all three stakeholder types. Furthermore, we provide the validated interview guide, which can assist future research in determining the explanation preferences of different stakeholder types. 

\end{abstract}

\section{Introduction}
Within the emerging field of explainable artificial intelligence (XAI), a substantial amount of research has attempted to make the inner workings of AI models more transparent \cite{hagras2018toward,mei2023users}. While such information can assist developers in understanding their model (e.g., by allowing the detection of bugs and biases, understanding feature importance), it is often complicated and requires considerable a priori knowledge of AI to interpret. However, the use of AI has become commonplace in user-controlled environments, such as the recommender systems used by different commercial platforms (e.g., YouTube, TikTok, Amazon). In such environments, explanations cannot assume AI knowledge, as the majority of explainees are lay users. Moreover, different types of users interact with such systems - the stakeholders. These stakeholders consist of every individual or group who affects, or is affected by, the delivery of recommendations to users \cite{abdollahpouri2020multistakeholder}. Stakeholders can be strongly diverse, coming from different backgrounds and having distinct expertise. As such, the way in which an explanation is conveyed to each stakeholder individually should be fine-tuned to their specific needs. 

One field where such fine-tuned explanations are especially crucial is recruitment. Recruitment is inherently a multi-stakeholder domain, as users (candidates) need to be linked to vacancies (provided by companies) by recruiters. These three main stakeholders all rely on the same recommendations but can require widely different explanations. For example, telling a candidate that a vacancy is relevant for them as it comes with a high salary can be an acceptable explanation. However, the same explanation will be useless for the company, as that salary will be provided to every other potential candidate. Furthermore, a candidate and a recruiter might only look at a handful of recommendations per session, while a company could receive hundreds of applicants for a single vacancy. Therefore, the explanation requirements of each stakeholder are unique and require a tailored design. 

This paper attempts to determine the explanation preferences of the stakeholders of a job recommender system: job seekers, companies, and recruiters. This is done through the execution of a co-design study, which allows stakeholder representatives to manually indicate how they prefer an explanation to be presented to them. Therefore, this research aims to answer the following research question: 

\begin{description}
    \item[RQ:] \textit{What are the explanation preferences of recruiters, candidates, and company representatives for job recommender systems?}
    \newline
\end{description}

Our results show interesting differences in the preferences of the different stakeholders. Regarding the preferred types of explanations, \textit{candidates} preferred brief written explanations, as their main interest is to be able to quickly judge the potential matches proposed by the system. On the contrary, company's \textit{hiring managers} preferred visual, graph-based explanations, as these allow a comprehensive overview at a glance. Finally, \textit{recruiters} preferred more exhaustive textual explanations, as those provided them with more talking points to convince both parties of the match. These results allow us to provide design guidelines for an interface that fulfills the requirements of all three stakeholder types. Furthermore, the co-design study allowed us to validate and improve the used interview guide. 

\section{Related work}
Within the field of explainable AI, there is no single agreed-upon method to provide explanations \cite{arya2019one}. Different use cases require different approaches, each with their own strengths and weaknesses. 

One of the most common methods of providing explanations is through text \cite{poli2021generation,cambria2023survey}. Textual explanations consist of brief sections of text that explain the rationale of the XAI model. Such texts often contain information on the impact different features had on the prediction and how those features interacted with each other. There are multiple ways to generate such texts, e.g., through the use of large language models (LLMs) \cite{menon2022visual} or predefined templates \cite{wrede2022linguistic}. 

Another popular approach is the use of feature attribution maps: visualizations that show the importance of different features to the prediction \cite{palacio2021xai}. Such maps can take different forms, depending on the specific task and data involved. When using tabular data, bar charts are often used to show the contribution of each different feature type to the prediction. When using multi-dimensional data, such as images or time series, are used, heatmaps can provide an overview of the importance of the different dimensions interacting with each other \cite{fauvel2021xcm}. 

A further explanation type that has been gaining popularity recently, is the knowledge graph-based explanation \cite{tiddi2022knowledge}. These explanations depend on the connections within a knowledge graph to explain the rationale behind a prediction. This is usually done by highlighting important nodes and edges within the graph, which provide `paths' from the subject to the recommended item, accompanied by their importance to the model's prediction \cite{wang2019kgat}.

\subsection{Challenges in multi-stakeholder explainability}
In multi-stakeholder environments, explanations need to meet additional requirements \cite{abdollahpouri2020multistakeholder}. An explanation that is sufficient for a developer, is not necessarily understandable for a user or provider, and vice versa \cite{szymanski2021visual}. There are multiple strategies to deal with this discrepancy, each with its own strengths and weaknesses. The most obvious solution is to create individual explanations for the different stakeholders \cite{yildirim2021bideepfm}. Although this leads to the most fine-tuned explanations, it introduces an additional layer of complexity to the system as a whole. Another approach would be to simply use a single explanation, but to present it differently based on the stakeholders' level of expertise \cite{abdollahpouri2020multistakeholder}. Unfortunately, it can be difficult to incorporate the different stakeholder perspectives simultaneously - some facts could be confidential or sensitive for a specific stakeholder, making it challenging to incorporate them in the explanation, even when they are relevant. Similarly, a highly specific overview of how the model came to the prediction might be useful for a developer, but will be too confusing for a lay user or provider. 

\subsection{Explainability in job recommender systems}
Explaining reciprocal recommendations, such as job recommendations, tends to be more difficult than standard recommendations, as the preferences of both parties need to be considered. Kleinerman et al. \cite{kleinerman2018providing} looked at explainability for recommender systems in online dating and found that explanations that consider both parties outperform one-sided explanations in high-cost scenarios (such as recruitment). In particular, explanations based on specific feature values are useful, although only a few features should be included to prevent information overload. In high-cost scenarios, explanations should not stay limited to personal preferences (e.g. `you should apply for this job because you want a company that has X attributes'), but should also incorporate an explanation of why the other party is likely to agree (e.g. `they are likely to accept you, because they are looking for a candidate with Y skills'). 

In job recommender systems (JRSs) specifically, explainability has largely gone unexplored. While some previous work has incorporated some degree of explainability within their JRSs, the explanations are often limited and seem to have been included as an afterthought \cite{le2019towards,Upadhyay2021,yildirim2021bideepfm}. Even when explainability has been included, authors usually fail to consider all stakeholders, tailoring the explanations to developers only. Furthermore, explanations are often solely evaluated anecdotally, leaving their quality up for debate \cite{nauta2022anecdotal}. One could argue that reciprocal, easy-to-understand explainability should be at the core of the models' design in a high-risk, high-impact domain such as recruitment. Where previous research mainly falls short, is in the understandability of their explanations: while their models can technically explain some part of their predictions, the explanations tend to be unintuitive and/or limited, either staying too vague \cite{le2019towards,Upadhyay2021} or being hard to understand \cite{yildirim2021bideepfm}. In previous work, we found that, when dealing with users with limited AI knowledge, such as recruiters, job seekers, and most company representatives, having clear, straightforward explanations is crucial \cite{schellingerhout2022explainable,szymanski2021visual}. To accomplish this, structured requirements engineering needs to be conducted in order to determine the preferences of all stakeholders, after which explainable JRSs will need to be designed with those requirements as a starting point. 

\subsection{Determining stakeholder preferences}
In order to determine the explanation preferences of different stakeholders, their requirements, struggles, and level of expertise need to be documented. To accomplish this, multiple approaches exist; for example, whenever the preferences of a stakeholder are already largely known (e.g., through previous research) questionnaires can be used in combination with different alterations of some explanation type \cite{szymanski2021visual}. The results of these questionnaires could then be used to `fine-tune' the already-known explanation type to better fit the exact stakeholders. However, within job recommendation, stakeholder preferences (beyond candidates) are mostly unknown \cite{de2021job}. Therefore, it is better to determine the stakeholder preferences from the ground up, allowing them to assist in shaping the explanations themselves. Thus, (semi-structured) interviews are highly useful, as they give stakeholders the freedom to indicate their exact preferences and requirements regarding explanations \cite{longhurst2003semi}. 

\subsection{Contributions}
Explainability within multi-stakeholder environments has largely gone unexplored. Research that has touched upon this topic, has often stuck to offline methods of evaluation, which fall short in high-impact domains, such as recruitment and healthcare. Therefore, this paper aims to lay the foundation for future research on explainable multi-stakeholder recommendation. We do so firstly by providing a validated interview guide that can be used to extract the explanation preferences of different stakeholder types. Furthermore, we extend the current literature on explainable job recommender systems by not just focusing on a single stakeholder, but providing guidelines on how explanations should be designed for all stakeholders involved. 
\section{Methodology}

In order to discover the preferences of different stakeholders, semi-structured interviews were conducted using example explanations \cite{garcia2018mixed}. During these semi-structured interviews, the participants were asked to answer substantive questions based on the provided explanations, as well as questions to indicate what aspects of different explanations they prefer. These substantive questions were used to gauge their understanding of the explanations. This is important, as previous research found that preference and understanding do not necessarily go hand in hand \cite{szymanski2021visual}. In our study, we are interested in particular in highlighting the specific explanation preferences of the specific stakeholders. Hence, we decompose our main research question into the following three sub-questions:

\begin{description}
    \item[SQ1:] What type of explanation is most suited for the different stakeholders?
    \item[SQ2:] What aspects of explanations make them more suited for each stakeholder?
    \item[SQ3:] In what way can different explanation types be combined to make them useful for each stakeholder?
\end{description}

\subsection{Hypotheses}

In this study, we consider three different explanation types (see Section \ref{sec:explanation_types}): (i) graph-based explanations; (ii) Textual explanations; and (iii) Feature attribution explanations. While the graph-based explanations will most likely be best suited for individuals with a fair amount of prior AI knowledge, the general lay users will probably gravitate towards the textual explanations, as those are both expressive and fairly easy to process \cite{purificato2021evaluating}. Considering the graph-based explanations contain the most information, but are expected to be the hardest to read, and the opposite holds for the feature attributions, the textual explanations are likely to strike a good balance between the two. These considerations lead us to formulate two hypotheses related to the \textbf{SQ1}:

\begin{itemize}
    \item \textbf{H1a}: \textit{The graph-based explanation will be best suited for individuals with prior AI knowledge.}
    \item \textbf{H1b}: \textit{The textual explanations will be best suited for individuals without prior AI knowledge.}    
\end{itemize}

Furthermore, we considered that feature attribution maps are usually the easiest and fastest way to get an overview of the model's rationale, but at the same time, they have a fairly limited extent \cite{szymanski2021visual}. The textual explanation will be more complex and take more time to process, but will provide a more comprehensive explanation in return. Lastly, the graph-based explanations will take the longest to process and might be difficult to interpret by themselves, but will contain the most complete explanation as a result. We then expect differences among the stakeholders, and formulate the following hypothesis related to the \textbf{SQ2}:

\begin{itemize}
    \item \textbf{H2}: \textit{The different stakeholders (candidates, companies, and recruiters) will have different preferences related to the explanation types.}   
\end{itemize}

Finally, we considered that explanations consisting of a single type may be either too limited in their content, or too difficult to interpret. This problem can be addressed by incorporating aspects from different types into a single explanation type \cite{szymanski2021visual}. For example, textual explanations can help in assisting the stakeholders in how to read the graph-based explanation. Furthermore, the feature attribution map can be useful when the stakeholder prefers to get a good (albeit limited) overview at a glance \cite{schellingerhout2022explainable}. We further hypothesize then that also regarding the preferences in terms of combining basic explanations into hybrid strategies, the stakeholders will have differences. Hence, we formulated the following hypothesis related to the \textbf{SQ3}:

\begin{itemize}
    \item \textbf{H3}: \textit{The different stakeholders (candidates, companies, and recruiters) will have different preferences on how to combine explanation types to obtain a hybrid explanation.}

\end{itemize}

\subsection{Semi-structured interview guide}
A comprehensive guide was created to conduct the semi-structured interviews (\cref{app:interview_guide}). However, this guide is susceptible to possible biases, ambiguities, incorrect assumptions about prior knowledge, etc. Thus, during the interviews, we dedicated time specifically to determining the quality of the questions in order to update, and eventually validate them. The questions in the interview guide were based upon previous works \cite{chen2005trust,cramer2008effects,kleinerman2018providing,pu2011user}, but required validation for a multi-stakeholder scenario. In addition to validating the interview guide, the interviews also allowed the stakeholders to co-design the explanation representations to fit their needs. The interviews were conducted with a small sample of the different stakeholders ($n = 2$ for each stakeholder type) to verify the adequacy of the explanations and the guide for each group individually. Considering the fact that each participant was interviewed three times (once for each explanation type), we collected a large amount of data per participant. Due to the richness of this data, the relatively small sample size still allowed us to perform an in-depth analysis for each stakeholder type. Previous works also indicates that a small sample can be sufficient in qualitative analysis, as long as the data itself is of high enough quality \cite{dworkin2012sample,morse2000determining}. Note that the user study was approved by the ethical committee of our institution.\footnote{Ethical Review Committee Inner City faculties (Maastricht University)}


\subsection{Data \& model}
The explanations used for the interviews were generated using a job recommendation dataset provided by Zhaopin.\footnote{\url{https://tianchi.aliyun.com/dataset/31623/}} This dataset contains information on 4.78 million vacancies, 4.5 thousand candidates, and 700 thousand recorded interactions between the two. For candidates, the dataset stores information such as their degree(s), location, current and desired industry, skill set, and experience. For vacancies, features such as the job description, job title, salary, (sub-)type, required education, location, and starting date were recorded. The interactions between candidates and vacancies consist of four stages: no interaction, browsed, delivered, and satisfied. Considering the data in the dataset was exclusively in Mandarin Chinese, all unstructured data was automatically translated to English using deep-translator.\footnote{\url{https://pypi.org/project/deep-translator/}}

These three different tables were combined together into a single knowledge graph, wherein candidates, vacancies, and categorical features formed the set of nodes. The edges consisted of relations between these nodes, such as candidates interacting with vacancies or vacancies being of a specific (sub-)type. This single, large knowledge graph was then converted to individual sub-graphs between candidates and vacancies that had interacted (positives), and between those who had not interacted (negatives) using the k random walks algorithm \cite{lovasz1993random}. Each of these sub-graphs therefore could be given a score from 0 (no interaction), to 3 (satisfied), which allowed us to treat the task as a ranking problem based on sub-graph prediction.

The explainable model was based on the graph attention network (GAT) \cite{velivckovic2017graph}, implemented using PyTorch geometric.\footnote{\url{https://www.pyg.org/}} Considering performance was not the goal of this research, we opted for a straightforward architecture consisting of a single GATv2Conv-layer, followed by two separate GATv2Conv-layers - one for the company-side prediction and explanation, and one for the candidate-side prediction and explanation. Both these layers were followed by their own respective fully-connected layer, which provided the company- and candidate-side prediction score. The harmonic mean of these two scores was then considered the final `matching score'. Optimization was done using the Adam optimizer \cite{kingma2014adam} (learning rate $= 1 * 10^{-3}$) with LambdaRank loss based on normalized Discounted Cumulative Gain @ 10 (nDCG@10) \cite{burges2006learning}. Hyperparameter tuning was done using grid search going over different configurations of hyperparameters \cite{liashchynskyi2019grid}. Since our aim was not to get state-of-the-art performance, the number of different configurations tested was fairly limited. Even so, the optimal configuration led to an nDCG@10 of 0.2638 (\cref{app:hyperparameters}).

Considering the goal of our research was not to evaluate the explanation quality of the specific model, but rather to investigate stakeholder preferences in general, the examples used during the interviews were manually selected based on the following criteria: graph size, perceived sensibility, and accessibility of the industry for evaluation. By sticking to seemingly sensible explanations that did not require knowledge of the specific industry at hand, we aimed to make the stakeholders' evaluation dependent solely on the representation of the explanation, rather than the quality of the model's explanations in general. 

\subsection{Explanation types}
\label{sec:explanation_types}
\begin{figure}[th]
    \caption{An example of knowledge graph paths being used as an explanation for a recommendation.}
    \begin{subfigure}[b]{0.5\linewidth}
         \centering
         \includegraphics[width=\textwidth]{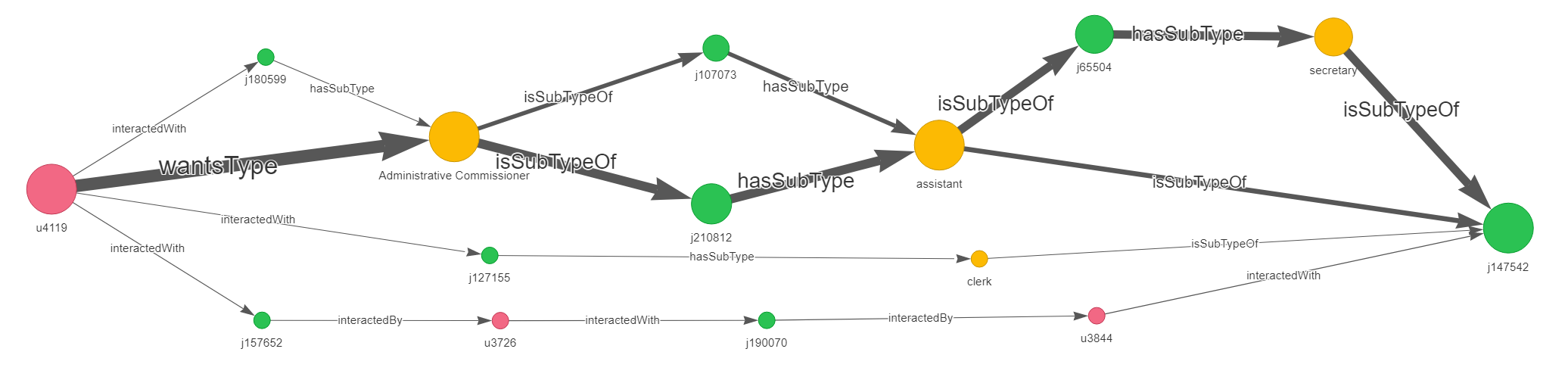}
         \caption{Candidate-side graph}
         \label{fig:cand_side}
     \end{subfigure}
     \hfill
     \begin{subfigure}[b]{0.5\linewidth}
         \centering
         \includegraphics[width=\textwidth]{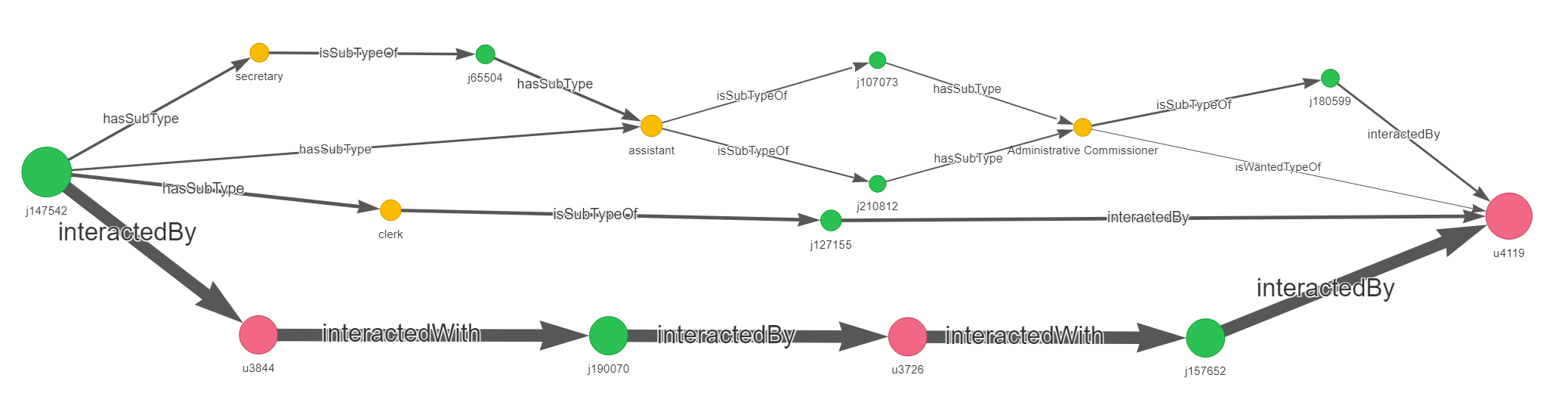}
         \caption{Company-side graph}
         \label{fig:comp_side}
     \end{subfigure}
    \label{fig:paths}
\end{figure}

The explanation types that were examined in this study were the following:

\begin{description}
    \item[Graph:] a visualization of paths in a knowledge graph. In our case, this consists of (a sub-set of) the paths within the candidate-vacancy sub-graph, weighted by the importance ascribed to them by the model (\cref{fig:paths});
    \item[Textual:] a short text that explains which features contributed to the recommendation in what way. The textual explanations are generated using a large language model (LLM) (in this case, ChatGPT February 13 version \cite{openai2022chatgpt}), which is given the full graph explanation as input, and tasked to summarize it in an easy-to-read way (\cref{fig:textual}); 
    \item[Feature attribution:] a visualization (such as a bar chart) that shows which features were most important to the model when creating the explanation (\cref{fig:attribution}). This bar chart is also based on the paths within the knowledge graph - the sizes of the bars are calculated using the sum of incoming edge weights, similar to PageRank \cite{bianchini2005inside}.
\end{description}

\newpage

\begingroup{
    \captionsetup{type=figure}
    \captionof{figure}{An example of a textual explanation used as an explanation for a recommendation}
    \fbox{\begin{minipage}{0.9\linewidth}
    The XAI model has analyzed various connections between jobs and users to determine if a particular user (\textcolor{xpink}{user 4119}) would be a good fit for a specific job (\textcolor{xgreen}{job 147542}). The model looked at the relationships between different jobs and users, as well as the importance of these relationships, to make its prediction.
    
    \vspace{0.1in}
    
    In this case, the model found that \textcolor{xpink}{user 4119} has a strong connection to the role of \textcolor{xyellow}{Administrative Commissioner}, and this connection is considered to be very important for explaining why \textcolor{xpink}{user 4119} would be a good match for \textcolor{xgreen}{job 147542}. Additionally, the model found that \textcolor{xgreen}{job 147542} has a connection to the role of \textcolor{xyellow}{secretary}, which is also considered important. The model also found that the \textcolor{xyellow}{Administrative Commissioner} role has a connection to the \textcolor{xyellow}{assistant} role, which in turn has a connection to the \textcolor{xyellow}{secretary} role and \textcolor{xgreen}{job 147542}.
    
    \vspace{0.1in}
    
    In summary, the XAI model determined that \textcolor{xpink}{user 4119} would be a good fit for \textcolor{xgreen}{job 147542} based on the strong connection between \textcolor{xpink}{user 4119} and the \textcolor{xyellow}{Administrative Commissioner} role, as well as the connections between the \textcolor{xyellow}{Administrative Commissioner} role, the \textcolor{xyellow}{assistant} role, the \textcolor{xyellow}{secretary} role, and \textcolor{xgreen}{job 147542}.
    \end{minipage}}
    \label{fig:textual}
}

\begin{figure}[bht]
    \caption{An example of a feature attribution map used as an explanation for a recommendation.}
    \centering
    \includegraphics[width=\linewidth]{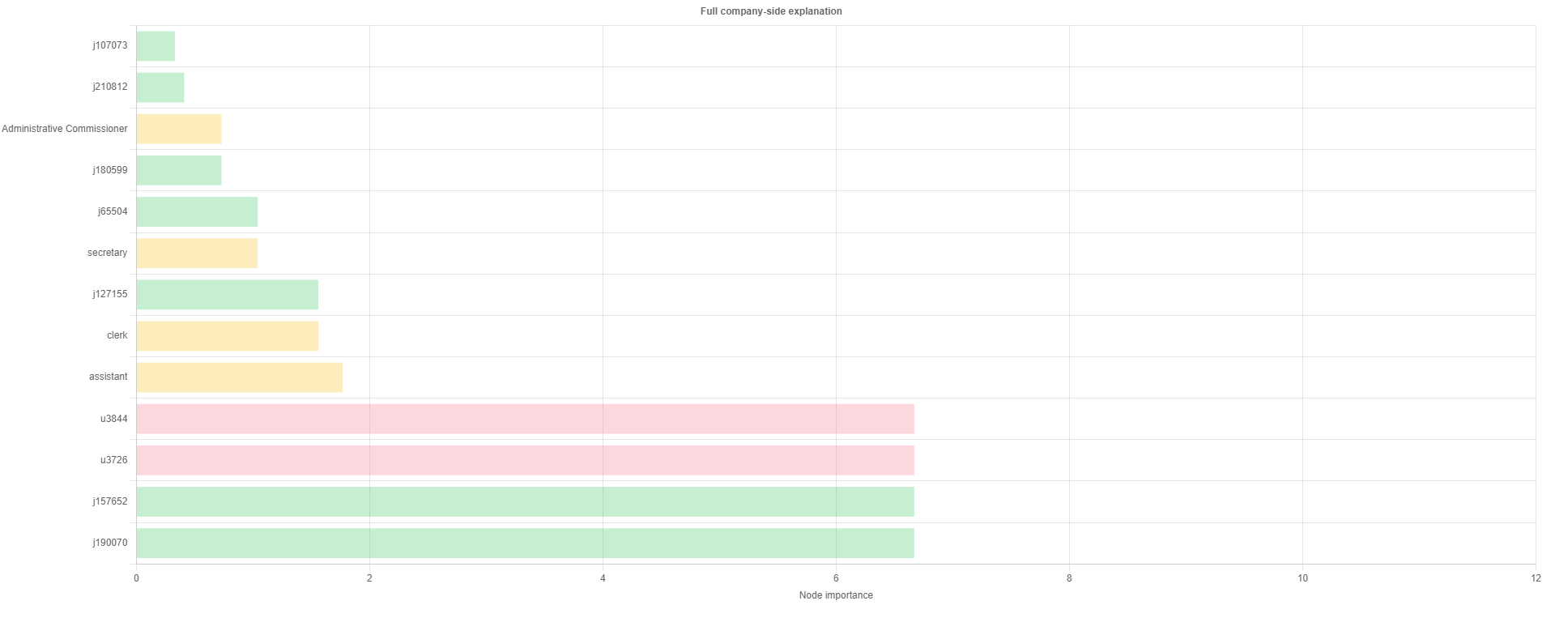}
    \label{fig:attribution}
\end{figure}


\subsection{Analysis}
The answers provided by the participants were analyzed using grounded theory (GT) \cite{walker2006grounded} using \textit{Atlas.ti}\footnote{\url{https://atlasti.com/}}. This process was done separately for each stakeholder type, in order to create distinct results for each type. We started by assigning manually-generated codes to individual statements made by the participants (open coding). These codes were then grouped together into distinct categories using axial coding to provide a higher-level overview of the complaints and preferences of each stakeholder type. Lastly, selective coding was conducted to combine each stakeholder type's categories into a single theory. The higher-level categories, as well as the theories, were then used to improve the prototypical explanation. 

\subsection{Participant demographics}
The participants were recruited through personal connections in collaboration with Randstad Groep Nederland\footnote{\url{https://www.randstad.nl/}}, the largest recruitment agency in the Netherlands. The sample consisted of 4 women and 2 men, of various ages ($\mu = 39.2, SD = 11.3, l=23, h=53$), with various backgrounds (tech, finance, healthcare, marketing, etc.). The participants had largely different levels of expertise w.r.t. AI, ranging from no knowledge whatsoever to a Bachelor's degree in a related field. Each interview took approximately one hour, and candidates were paid \euro{11,50} for their time. 
\section{Results}
Based on the answers given by the different stakeholder representatives, the interview guide has been updated, and the preliminary preferences of each stakeholder type have been determined. The full transcripts of each interview are available on \href{https://github.com/Roan-Schellingerhout/JRS_explanations}{GitHub}. 

\subsection{Interview guide}

The interview guide (\cref{tab:interview_guide}) has been validated based on responses from the participants. While the interview guide was largely proven to be adequate for determining the explanation preferences of different stakeholders, some changes have been made based on the feedback we received. 

Firstly, we added an additional question to the section on \textbf{usefulness}: `\textit{how could you see yourself using the explanation in your daily work/task?}' Even when some of the participants could see that an explanation was sensible, or could be helpful in making a decision, they mentioned that they would personally stick to using another approach (e.g., a different type of explanation, or doing things manually). Although this is likely to come up using the current interview guide already, we decided to also explicitly ask the question - after all, the goal of creating an explainable model is that users end up actually using the explanations to assist them in their decision-making. 

We additionally added a new question to the \textbf{correct interpretation} section of the guide: ``\textit{how would you put the model's explanation into your own words?}'' While the participants often managed to quickly identify the most/least important features and components of the explanation, that did not necessarily indicate they properly understood the entire rationale. For example, some participants correctly identified the importance of the different job types to the explanation, but they could not properly connect all the dots, causing them to be unable to view the explanation as a single whole. By explicitly asking them to define the explanation in their own words, it becomes clear whether they adequately understand the entire explanation, or are still grasping at straws.

Lastly, we changed one of the questions in the section on \textbf{transparency}: ``can you think of anything that would further improve your understanding?'' has been changed to ``what information is missing that could allow you to get a better understanding of the model's recommendation?''. The previous phrasing of the question was too general, making it difficult for participants to answer it on the spot. By directly asking them for information that is missing, it should be easier for them to come up with an answer, albeit a more indirect one.

The updated, validated interview guide can be seen in \cref{tab:interview_guide_updated}.

\begin{table*}[!t]
\captionsetup{width=\textwidth}
\caption{The validated, updated interview guide.}
\label{tab:interview_guide_updated}
\scriptsize
\begin{adjustwidth}{-2.5in}{-2.5in}
\centering
\begin{tabularx}{1\textwidth}{@{}XX>{\raggedright\arraybackslash}p{3.75cm}>{\raggedright\arraybackslash}p{4.5cm}@{}}
\toprule
\textbf{Evaluation Objective} & \textbf{Objective Description} & \textbf{Questions} & \textbf{Probing questions} \\ \midrule

1. Correct interpretation     & To assess whether or not the stakeholder can correctly interpret the explanation. & \begin{enumerate} \item[1.1] What information/features do you think were most important for this prediction? \item[1.2] What was the least important? \item[1.3] How would you put the model's explanation into your own words? \end{enumerate} &         \begin{enumerate} \item[1.1.1] What did you look at to come to that conclusion? \end{enumerate} \\ \midrule

2. Transparency & To determine the explanation's effect on understanding the model's inner workings.                & \begin{enumerate} \item[2.1] Does the explanation help you comprehend why the system gave the recommendation? \end{enumerate} & \begin{enumerate} \item[2.1.1] What components help you specifically? \item[2.1.2] What information is missing that could allow you to get a better understanding of the model's recommendation \end{enumerate} \\ \midrule

3. Usefulness   & To evaluate how useful the explanations are considered to be.                    &     \begin{enumerate} \item[3.1] Does the explanation make sense to you? \item[3.2] Does the explanation help you make a decision? \item[3.3] How could you see yourself using the explanation in your daily work/task? \end{enumerate} & \begin{enumerate} \item[3.1.1] What do you consider sensible (e.g., focus on specific features)? \item[3.1.2] What do you consider insensible? \item[3.2.1] Would you prefer a model with explanations over one without? \end{enumerate} \\ \midrule

4. Trust        & To gauge the explanation's impact on the model's trustworthiness.      &     \begin{enumerate} \item[4.1] Do you think the prediction made by the model is reliable? \item[4.2] If this recommendation was made for you, would you trust the model to have made the right decision? \end{enumerate} & \begin{enumerate} \item[4.2.1] Anything specific that makes you say that (e.g., something makes no sense, or is very similar to how you look at things)? \end{enumerate} \\ \midrule

5. Preference   & To figure out the personal preference of the stakeholder.                        &     \begin{enumerate} \item[5.1] What would you like to see added to the current explanation? \item[5.2] What would you consider to be redundant within this explanation? \end{enumerate} &  \begin{enumerate} \item[5.1.1] Any specific information that is missing? \item[5.1.2] Any functionality that could be useful? \item[5.2.1] Anything that should be removed? \item[5.2.2] Or be made less prevalent?\end{enumerate}             \\ \bottomrule
\end{tabularx}
\end{adjustwidth}
\end{table*}

\subsection{Stakeholder preferences.}
Each of the transcripts has been analyzed, and the analyses have been grouped based on stakeholder type. An overview of the generated codes, categories, theories, and relevant quotes can be found in \cref{app:GT}.

\subsubsection{Candidates.}
In line with our hypothesis, the textual explanation was well-received by the candidates. Although the candidates did receive the explanations favorably, they indicated some issues that should be addressed. For one, the specific language used in the explanation made it more difficult for the candidates to parse it correctly. Candidates also sometimes wound up losing track of the essence of the explanation, specifically when multiple trains of thought were addressed in a single paragraph, or whenever points were reiterated multiple times (\textit{``it's a bit more clear, but I don't know... I still can't follow it completely. I find it very hard to read"}, P2, Q1.1). However, the candidates still managed to correctly identify the main arguments on which the recommendation was based. They did indicate that they would prefer to be able to evaluate the text at a glance, i.e., by putting crucial information clearly at the top of the text (`\textit{`Information like the city, and what the salary is, or things of the sorts, are currently not included"}, P1, Q4.2). They did not fully trust the model, but found it to be a nice `brain-storming partner', which could support them in their decision (\textit{``if I had any doubts, the explanation would take those away''}, P2, Q3.2), and provide them with some interesting vacancies to explore on their own. Furthermore, the explanation contained the full ID of candidates and vacancies relevant to the recommendation. Considering the candidates have no access to the actual database, these IDs turned out to be of little value, and actually overwhelmed the candidates. Additionally, while multiple different vacancies were mentioned in the explanation, these were not directly accessible to the candidates - something they considered to be unintuitive. 

The graph-based explanation turned out to be difficult to use for the candidates without receiving some additional help on how to interpret it. Especially the full, unsimplified version of the graph, in which all of the different paths were visible, turned out to be too overwhelming and complex to be useful (after being corrected on their interpretation: \textit{``Now I get that the thin lines were kind of like side tracks that weren't successful"}, P2, Q2.1). However, with some additional guidance on how it was structured, and by considering the simplified view of the graph, the candidates eventually correctly understood its content. Still, the amount of information contained within the graph was more confusing than helpful - for example, the types and values of the edges were constantly visible, meaning there was a lot of text present at all times. Considering the candidates did not necessarily understand the meaning of the edge types and values, they did not get any benefit from them. While the graph was indicated to give them a better understanding of the model's actual rationale, it also made them question the adequacy of the recommendation to an extent. Because vacancies different from the one being recommended were included in the graph as well, sometimes at locations seemingly `closer' to the candidate, they were unsure why \emph{those} vacancies were not recommended to them (\textit{``It feels quite strange, that the path passes through different relevant vacancies, but we just ignore those"}, P1, Q1.1). Therefore, providing candidates with information on why alternative vacancies were not recommended could be helpful. 

The current implementation of the feature attribution chart turned out to be close to useless for providing an explanation to the candidates. Since it ascribed importance to different vacancies and candidates (i.e., similar vacancies and candidates, which were included to allow for collaborative filtering \cite{su2009survey}), which they were not familiar with, it did not contribute to their understanding of the model's prediction. However, they did indicate how feature attribution could be made useful: for the bar chart to relate to their individual skills, so the candidates could understand which of their own skills were considered most important for the recommendation at a glance (\textit{``That it shows what is important for the match, for example, that your experience with Excel matters, but that your ability to be a truck driver wasn't important"}, P1, Q1.1). This would allow them to quickly verify and scrutinize recommendations. For example, if they saw a specific skill they possess was attributed a lot of importance by the model, even though they would not enjoy performing it as their job. Thus, the feature attribution chart should stick only to the personal, actionable features of the candidates. To still include previous vacancies fulfilled by the candidate, they could alternatively be grouped by job type, so that they could be represented as a single bar relating to their experience in that field (\textit{``the function types make sense to me, but the individual vacancies and candidates do not"}, P1, Q3.1).

\subsubsection{Recruiters.}
As was hypothesized, the recruiters found the textual explanations to be informative and useful. Although they found the texts to indicate some redundancy, as well as some tricky language, they considered them to be rather useful regardless. They immediately understood the main message of the text, but found that some information was reiterated too often (\textit{``It keeps beating around the bush with the same words"}, P3, Q2.1). The recruiters explicitly stated that they would not blindly trust the model, even when accompanied by a sensible explanation - they would always want to be able to manually verify its recommendation by reading the CV and vacancy text (\textit{``Very little is told about the candidate, and the vacancy, but I just have to trust that ... it would be nice to check if it's actually right"}, P3, Q2.1). However, as long as they considered that the explanation made sense, they would ``\textit{move that CV and vacancy to the top of their list}''. Furthermore, being able to quickly rule out specific candidates or vacancies was something they considered highly important. As a result, they strongly preferred to have the most crucial information, such as commuting time, and whether some minimum requirements were met, to be at the top of the explanation. The recruiters internally disagreed on how long the text should be (\textit{``it doesn't need to be brief. It's nice to have things to talk about"}, P3, Q5.2; \textit{``Three paragraphs of three sentences would be fine"} P4, Q5.1). One argument for having a longer text, was that it would provide the recruiter with more subject matter to discuss when trying to convince a candidate or company of the match's aptitude. On the other hand, having a shorter text that only focuses on the main arguments provided by the model would make it quicker for recruiters to compare different recommendations, after which they could come up with further arguments for the best one themselves. 

Although the recruiters managed to correctly interpret the graph, they found it to add little value for the most part (\textit{``I understand it, but it means little to me"}, P4, Q1.1). While they indicated that it could be useful in some specific scenarios, such as texts with a high level of complexity, or when having to support their final decision to a supervisor, they generally did not consider it to add any benefit compared to the textual explanation. Despite the fact that they developed a better understanding of the models' actual rationale, they doubted that they would use it much in their day-to-day tasks (\textit{``If I would have to use this for every vacancy, or every candidate, it would become a problem"}, P4, Q1.1). In the scenarios where they would consider it to be beneficial, they gravitated strongly toward the simplified version of the graph, considering the connections deemed unimportant by the model to be counterproductive in understanding the explanation. Only when a candidate or company would ask specifically about whether a specific skill, or past job, was taken into account, would they make use of the full version of the graph. 

Similarly to the candidates, the feature attribution chart in its current form did not assist the recruiters in correctly understanding the explanation. Again, though, did they indicate that a different type of bar chart could be useful in some scenarios. One use case for a bar chart the recruiters considered useful, was for it to be a central `hub' of sorts, where all possible vacancies for a candidate (and vice versa) were displayed, sorted by their matching score (\textit{``No, I would personally go for something like a top 10, for example"}, P3, Q2.1). This would allow the recruiters to quickly determine which potential matches are feasible enough to explore further. However, as an actual method of explaining the prediction, they indicated that the text, sometimes combined with the graph, would already be sufficient, causing the feature attribution chart to be largely irrelevant.

\subsubsection{Company representatives.}
As opposed to the candidates and recruiters, the company representatives were less positive about the textual explanation. Considering the company-side explanation contained a higher level of abstraction, it took the company representatives multiple iterations before they properly understood the explanation (\textit{``Now that I read it again, I see that it goes from person A, to B, to C."}, P5, Q3.2). They also found it difficult to take the explanation at face value, being wary of terms such as `relevant experience' - rather opting to manually verify whether the mentioned experience was actually relevant for the vacancy (\textit{``But judging how relevant the experience is for this vacancy, isn't possible based on this explanation"}, P5, Q1.2). Although a more detailed explanation could alleviate some of this hesitance, it would also lead to an even more complex explanation, possibly worsening the understandability further. 

On the other hand, the company representatives were considerably more positive about the graph explanation. Specifically, the simplified view of the graph allowed them to grasp the prediction at a glance. Where the textual explanation required some puzzling before the relations between different candidates, vacancies, and job types became clear, the company representatives quickly managed to detect the relevant relations in the graph (\textit{``This adds everything I need ... For me it's simply a matter of checking why the model made its decision - that being the line at the bottom, and that would be all"}, P6, Q2.1). As a result, the graph explanation also improved their trust in the model; one recruiter even mentioned that, given a high-enough performance of the model, they would simply use the simplified graph as a sanity check, fully trusting the model if the explanation seemed somewhat reasonable (\textit{``If the model does what it says it does, I would simply trust it"}, P6, Q3.2). 

The feature attribution map was again received poorly. The company representatives indicated that it did not help them get a better understanding of the model's reasoning compared to the graph (\textit{``It's usable, but it doesn't clarify why we ended up with the recommended candidate"}, P5, Q2.1). However, the company representatives did consider the feature attribution chart to be useful in a different scenario - to verify the model paid no attention to irrelevant details. With some tweaks, such as changing vacancy IDs into actual titles, the bar chart would allow company representatives to make sure the model did not pay attention to something that the company determined to be irrelevant to the position. Furthermore, an aggregated version of the feature attribution chart, which displays which types of features were considered most by the model, could help the company representatives parse the textual explanation more easily, allowing them to direct most of their attention to the more important information (\textit{``This is what I was looking for while reading the text. I tried to determine these values in my head, but I kept getting distracted"}, P5, Q2.1).
\section{Discussion}
We discuss our results in relation to the three sub-research questions, for each type of stakeholder: which type of explanation is most suited, what makes these explanations most suitable, and can different explanation types be meaningfully combined. We discuss each research question in turn. 

\subsection{SQ1: What type of explanation is most suited for the different stakeholders?}

When analyzing the results regarding preferences for the different stakeholders, we notice that \textit{candidates} prefer short, clearly structured, straightforward texts, which allow them to quickly browse and judge the vacancies. These texts should include what they consider the most important information: features like travel distance, salary, and minimum requirements. This information should be central and easy to identify, preferably in bullets. \textit{Recruiters} also prefer texts, but disagree on the amount of text that is required. Thus, a short text, that centrally mentions potential `deal-breakers' and gives the main few arguments to motivate why a match was made should be the default. However, recruiters also prefer to have access to a more exhaustive text, which can provide them with more material that could be used to convince both parties (candidate and company) to agree with the match. On the contrary, \textit{company representatives} prefer graph-based explanations, as those assist them in quickly getting an overview of even more complex explanations at a glance. Such a representation also allows them to quickly scan different information, while reading would require more time and effort. These results are somewhat in line with our hypotheses H1a and H1b, but not entirely. While one of the candidates did have a lot of knowledge of AI, she still preferred the textual explanation. On the other hand, both company representatives did not have a strong background in AI, but preferred the graph-based explanations. We argue that it is not the AI knowledge per se that makes the graph preferable, but the amount of experience with, and affinity towards, reading graphs and charts.

\subsection{SQ2: What aspects of explanations make them more suited for each stakeholder?}

We also looked more specifically at the motivation for why certain stakeholders prefer certain explanations. For \textit{candidates}, the textual explanations were largely preferred due to their simplicity, and because they felt more `personal'. In particular, they preferred texts using simple language, that is clearly structured and not longer than a few short paragraphs. \textit{Recruiters} also preferred the textual explanations, due to their simplicity compared to the graph- and feature-based explanations. In particular, they struggled to interpret the visualizations and felt quite overwhelmed due to their `math-heavy' nature. Furthermore, the text directly assists them in their day-to-day tasks, as they can almost use some of the paragraphs verbatim to try and convince companies and candidates of the adequacy of a match. Finally, for \textit{company representatives}, the graph-based explanations were preferred, largely due to their ability to make more complex, high-level connections, within the data visible at a glance. Within the textual explanations, it became difficult for them to figure out the full line of reasoning of the model, due to there being a lot of `steps' from the vacancy to the candidate, which made it hard to process. The bar chart also made the text more accessible, but the graph-based explanation was considered a better option.

\subsection{SQ3: In what way can different explanation types be combined to make them useful for each stakeholder?}

Finally, we evaluated the stakeholders' preferences in terms of hybrid explanations, indicating how to combine different explanations together. Our results highlight how the feature-based explanation was poorly received by all stakeholders; however, they also indicated it to have potential, in case it is used to support one of the other explanation types. The unanimous aversion to the feature-based explanation was likely due to their failure to find a niche, either being too general to be useful in the simplified version, or too specific (and thus overwhelming) in the full version. For both \textit{candidates and recruiters}, the textual explanations should be the center of the explanation, by default in its simplified form. The user should then have the possibility to access additional information, using a toggle to get a more descriptive version of the text. Considering the difficulty of conveying relative importance levels within a text, the feature attribution map can be linked (i.e., through matching color coding) to clarify the text. The graph-based explanation can then be used as an optional addition, in case the text itself is not clear enough, or if the user wants even more evidence for a specific suggestion. 

On the contrary, \textit{company representatives} prefer the simplified graph to be central, supported by a textual explanation. Within the graph, most details (e.g., exact values) should be made optional, so that it is not overwhelming: they mainly want to be able to quickly parse the most critical paths in the graph, and only look at details when necessary. Additionally, a bar chart indicating how important different feature types were, could be used to complement the text, in order to help them focus their attention on the paragraphs touching on those feature types.

\subsection{Limitations and future work}
A key limitation to acknowledge is the relatively small sample size we used. Considering we only interviewed two individuals from each stakeholder type, it is possible that some of our results are based on their personal biases, which may not be representative of the entire population. We attempted to minimize these biases through careful selection of participants, making sure to include individuals from different backgrounds, both in terms of their expertise and personal characteristics. This limited sample did allow us to focus on the quality of the data we collected; due to the limited number of participants, it was feasible to interview them for longer periods of time. Considering the large amount of data gathered through the interviews, we believe that this limitation does not deteriorate the quality of the findings. Furthermore, the aim of this paper was to lay the groundwork for future research on explainable job recommender systems through the creation of a reusable interview guide, as well as determining general stakeholder preferences and differences. I.e., our aim was not to conclusively determine the exact, ultimate preferences of the stakeholders, but rather to allow for future research to have a solid foundation for more specific research. Regardless, future research should aim to make use of the validated interview guide with a larger sample. By making use of the guidelines we provided on how to represent automatically-generated explanations, it should be possible to design a single, hybrid explanation, that can be evaluated more quickly. As a result, interviews will take less time, making it more practical to use a larger sample size. 

Furthermore, we acknowledge room to improve the model we used to generate explanations, and that it may therefore not have generated the most sensible explanations. To counteract this behavior, we manually selected explanations that seemed suitable for the interviews. 
Future work could therefore use the provided interview guide to evaluate and compare explanations generated using a number of different techniques (e.g., attention mechanisms, saliency, post hoc methods). Different model architectures could also be compared in order to determine which architectures generate better explanations (either for a specific evaluation objective, or as a whole).

Lastly, another venue for future research could be to evaluate the textual explanations generated through different means. Although the explanations generated by ChatGPT were sufficient for our study, comparing them to differently-generated explanations (e.g., by different LLMs, people with varying levels of expertise) could lead to interesting insights.

\subsection{Conclusion}
In this paper, we aimed to develop and validate an interview guide for determining the explanation preferences of different stakeholder types. Additionally, we aimed to establish guidelines for creating XAI-generated explanations for different stakeholders within the field of job recommendation. The interview guide was largely proven to be adequate for determining the preferences of different stakeholders; a few minor changes were made to it in order to attain more concrete responses from the participants. Through the use of the interview guide, we found that candidates prefer explanations to take the shape of a short, clearly-structured text, that centrally contains the most crucial information. Recruiters, on the other hand, also preferred textual explanations, but were less strict on it having to be brief - indicating that longer texts could be useful in some scenarios. Company representatives indicated a preference towards graph-based explanations, as those allowed them to get a comprehensive overview of even more complex explanations. 

\bibliography{sources}
\bibliographystyle{splncs04}

\section{Appendix}
\appendix
\section{Hyperparameter tuning}
\label{app:hyperparameters}

The optimal hyperparameter configuration we found, is the following: 

\begin{itemize}
    \item hidden\_dimensions=10
    \item output\_dimensions=100
    \item number\_of\_layers=2
    \item attention\_heads=5
    \item dp\_rate=0.01
    \item learning\_rate=0.001
    \item epochs=1
\end{itemize}

\noindent An overview of all configurations we tested, can be found on \href{https://github.com/Roan-Schellingerhout/JRS_explanations}{GitHub}

\section{Preliminary interview guide}
\label{app:interview_guide}

\begin{table*}[h]
\captionsetup{width=\textwidth}
\caption{The preliminary interview guide.}
\tiny
\begin{adjustwidth}{-2.5in}{-2.5in}
\centering
\begin{tabularx}{1\textwidth}{@{}XX>{\raggedright\arraybackslash}p{3.5cm}>{\raggedright\arraybackslash}p{4.5cm}@{}}
\toprule
\textbf{Evaluation Objective} & \textbf{Objective Description} & \textbf{Questions} & \textbf{Probing questions} \\ \midrule

1. Correct interpretation     & To assess whether or not the stakeholder can correctly interpret the explanation. & \begin{enumerate} \item[1.1] What information/features do you think were most important for this prediction? \item[1.2] What was the least important? \end{enumerate} &         \begin{enumerate} \item[1.1.1] What did you look at to come to that conclusion? \end{enumerate} \\ \midrule

2. Transparency & To determine the explanation's effect on understanding the model's inner workings.                & \begin{enumerate} \item[2.1] Does the explanation help you comprehend why the system gave the recommendation? \end{enumerate} & \begin{enumerate} \item[2.1.1] What components help you specifically? \item[2.1.2] Can you think of anything that would further improve your understanding? \end{enumerate} \\ \midrule

3. Usefulness   & To evaluate how useful the explanations are considered to be.                    &     \begin{enumerate} \item[3.1] Does the explanation make sense to you? \item[3.2] Does the explanation help you make a decision? \end{enumerate} & \begin{enumerate} \item[3.1.1] What do you consider sensible (e.g., focus on specific features)? \item[3.1.2] What do you consider insensible? \item[3.2.1] Would you prefer a model with explanations over one without? \end{enumerate} \\ \midrule

4. Trust        & To gauge the explanation's impact on the model's trustworthiness.      &     \begin{enumerate} \item[4.1] Do you think the prediction made by the model is reliable? \item[4.2] If this recommendation was made for you, would you trust the model to have made the right decision? \end{enumerate} & \begin{enumerate} \item[4.2.1] Anything specific that makes you say that (e.g., something makes no sense, or is very similar to how you look at things)? \end{enumerate} \\ \midrule

5. Preference   & To figure out the personal preference of the stakeholder.                        &     \begin{enumerate} \item[5.1] What would you like to see added to the current explanation? \item[5.2] What would you consider to be redundant within this explanation? \end{enumerate} &  \begin{enumerate} \item[5.1.1] Any specific information that is missing? \item[5.1.2] Any functionality that could be useful? \item[5.2.1] Anything that should be removed? \item[5.2.2] Or be made less prevalent?\end{enumerate}             \\ \bottomrule
\end{tabularx}
\label{tab:interview_guide}
\end{adjustwidth}
\end{table*}

\newpage

\section{Grounded theory results}
\label{app:GT}

\subsection{Candidates}
\begin{table*}[]
\captionsetup{width=1\textwidth}
\caption{The quotes, open codes, and categories discovered by using grounded theory for the candidates' responses.}
\tiny
\begin{adjustwidth}{-2.5in}{-2.5in}
\centering
\begin{tabularx}{1\textwidth}{@{}XX>{\raggedright\arraybackslash}p{3.5cm}>{\raggedright\arraybackslash}p{4.5cm}@{}}
\toprule
\textbf{Quotes} &
  \textbf{Open codes} &
  \textbf{Category} \\ \midrule
``You should separate the recommended and supporting vacancies" &
  Different instances of the same group should be easily distinguishable &
  Don't mix different types of information \\ \cmidrule(r){1-1}
``I find it very difficult that the vacancies, candidates, and vacancy types or on the same axis ... I don't understand it anymore" &
  Having different feature types in the same bar chart is confusing &
   \\ \cmidrule(r){1-1}
``if this was important for me as a candidate, I would want to know" &
  The bar chart should refer only to personal information &
   \\ \midrule
``It could be that the system is missing some information about you ... like that you want to work from home ... which would allow you to cross it off the list immediately" &
  Make `deal-breakers' extremely clear &
  An explanation should very quickly allow for verification and scrutiny \\ \cmidrule(r){1-1}
``A candidate would not want to spend all of their time dissecting a graph" &
  Having to extract information carefully is a bother &
   \\ \cmidrule(r){1-1}
``I would definitely want to look at the vacancy" &
  Manual follow-up should be easy &
   \\ \midrule
``And that those small lines, for us, as people looking for a vacancy, are not very useful" &
  Only include supporting arguments &
  Make all non-crucial information optional \\ \cmidrule(r){1-1}
``Yes, because this is some pretty difficult use of language ... and those values are not clear to me at all to be honest" &
  Specific values lead to overwhelm &
   \\ \cmidrule(r){1-1}
``Now it's saying the same thing for the third time in a row already" &
  Only repeat information when summarizing &
   \\ \bottomrule

\end{tabularx}
\end{adjustwidth}
\label{tab:candidates_GT}
\end{table*}

\noindent \textbf{Theory (based on \cref{tab:candidates_GT}):} Candidates want to be able to determine whether or not a vacancy is relevant at a glance. To do so, the explanation needs to be brief and straight to the point. Once the candidate has found a potentially interesting vacancy, they should be able to explore the explanation in more detail. Considering their difficulty in parsing both the graph and feature attribution explanation, the textual explanation should always be central, with the other two merely functioning as further support. 

\newpage

\subsection{Recruiters}
\begin{table*}[]
\captionsetup{width=1\textwidth}
\caption{The quotes, open codes, and categories discovered by using grounded theory for the recruiters' responses.}
\tiny
\begin{adjustwidth}{-2.5in}{-2.5in}
\centering
\begin{tabularx}{1\textwidth}{@{}XX>{\raggedright\arraybackslash}p{3.5cm}>{\raggedright\arraybackslash}p{4.5cm}@{}}
\toprule
\textbf{Quotes} &
  \textbf{Open codes} &
  \textbf{Category} \\ \midrule
``It's nice when there's more text to talk about on the phone, as long as it's not the same thing over and over again" &
  A lot of text can help in having enough subject matter while talking to clients &
  The explanation should be useable as evidence while justifying a match to a client \\ \cmidrule(r){1-1}
``if you want to back up your decision during a meeting, where they expect reports and whatnot, it would be very nice" &
  The graph can provide a more `objective' explanation &
   \\ \cmidrule(r){1-1}
``if the first paragraph is about their skills, the second about their experience, and the third about their interests, a longer text would still be nice" &
  Each paragraph of the text should address a different aspect &
   \\ \midrule
``I don't think this is required to actually start calling; it's more of a convenience when you want to understand the reasoning" &
  Knowing the general rationale is enough to take action already &
  The exact details of the prediction are irrelevant most of the time \\ \cmidrule(r){1-1}
``There's a few things that are crucial when making a match ... and if those are not in order, I don't even need to see the prediction" &
  Possible points of contention should already have been considered &
   \\ \cmidrule(r){1-1}
``I don't want to know anything I don't need to know ... there's no use in that" &
  The simple version of explanations is usually sufficient &
   \\ \midrule
``you simply get told that this is the correct match ... and if you can look at the vacancy, you can check if it's correct" &
  Recruiters should be able to easily verify the model's claims &
  Recruiters should always feel like they have the final say \\ \cmidrule(r){1-1}
``I would never blindly set up a meeting; I would always want to speak to the candidate beforehand of course" &
  Recruiters first want to discuss the match with both sides before accepting it &
   \\ \bottomrule
   
\end{tabularx}
\end{adjustwidth}
\label{tab:recruiters_GT}   
\end{table*}

\noindent \textbf{Theory (based on \cref{tab:recruiters_GT}):} Recruiters prefer the model to act mainly as a supportive tool. This means that the strongest arguments the model puts forward should be front and center. This allows them to use the explanations when defending their decision, be it to their supervisor or a client. They will always want to manually verify the claims made by the model, but due to the explanation, they are likely to consider predicted matches before all else. The exact details of how the model came to its prediction will oftentimes be irrelevant, but are nice to have accessible in case additional evidence should be provided. 

\newpage

\subsection{Company representatives}
\begin{table*}[]
\captionsetup{width=1\textwidth}
\caption{The quotes, open codes, and categories discovered by using grounded theory for the company representatives' responses.}
\tiny
\begin{adjustwidth}{-2.5in}{-2.5in}
\centering
\begin{tabularx}{1\textwidth}{@{}XX>{\raggedright\arraybackslash}p{3.5cm}>{\raggedright\arraybackslash}p{4.5cm}@{}}

\toprule
\textbf{Quotes} &
  \textbf{Open codes} &
  \textbf{Category} \\ \midrule
``this is what I had, after reading the text four times, this path is generally what I had understood" &
  The textual explanation can require multiple iterations to become clear &
  Complex relations should still be easy to grasp \\ \cmidrule(r){1-1}
``at one point you understand how it works ... and then you won't even look at the text anymore, the graph will be all you need" &
  The graph is quick and easy to use once it's understood &
   \\ \cmidrule(r){1-1}
``the complex graph should be banned" &
  The general idea of the explanation should be clear at a glance &
   \\ \midrule
``that's why I would rather pick that one, over the recommended one, because that one seems closer to the vacancy." &
  Alternative candidates should also receive explanations &
  Exploring alternatives should be integrated in the system \\ \cmidrule(r){1-1}
``it could be a close call, you know? So then you can make your own assessment, and verify if the model got it right" &
  Having an overview of all possible candidates is useful to verify and scrutinize &
   \\ \cmidrule(r){1-1}
``if we want someone for 0.8 FTE, but their motivation letter says 0.6 FTE, it already becomes a no-go" &
  Human factors, such as candidates' motivation letters, are hard to integrate into a prediction &
   \\ \midrule
``if the model has been designed in such a manner that I know it has checked everything, there's no need for me to manually check everything as well" &
  Given high enough performance, the explanation merely becomes a sanity check &
  Explanations are mainly useful for surprising results \\ \cmidrule(r){1-1}
``you already have an expectation of what the outcome will be. You're only going to start interrogating the model once the prediction doesn't match your expectation" &
  Detail only matters when the recommendation is unintuitive &
   \\ \bottomrule
   
\end{tabularx}
\end{adjustwidth}
\label{tab:companies_GT}   
\end{table*}

\noindent \textbf{Theory (based on \cref{tab:companies_GT}):} Company representatives want the explanations to assist them as quickly as possible. Due to their generally higher level of experience in reading charts and graphs, the graph explanations actually help the most with this. However, even though the graph can give them an explanation at a glance, they still want to be able to explore further, in case the graph comes across as surprising or unintuitive. In such a scenario, they either want to study the explanation in more detail, e.g., through additionally reading the textual explanation, or they want to manually look into alternative candidates. The feature attribution map could easily be converted into a `hub' for them, where they can get an overview of alternative candidates for a vacancy.

\end{document}